\newcounter{myctr}

\documentclass{ws-raps}
\usepackage{url,multirow,color,graphicx,epsfig}

\begin{document}

\makeatletter
\def\@biblabel#1{[#1]}
\makeatother

\markboth{H.-L Shi, Z.-Q. Jiang and W.-X. Zhou}{Time-varying return predictability in the Chinese stock market}

%
\catchline{}{}{}{}{}
%

\title{Time-varying return predictability in the Chinese stock market}

\author{\footnotesize HUAI-LONG SHI}
\address{Department of Finance and Research Center for Econophysics,\\ East China University of Science and Technology,\\ 130 Meilong Road, Shanghai 200237, China}

\author{\footnotesize ZHI-QIANG JIANG}
\address{Department of Finance and Research Center for Econophysics,\\ East China University of Science and Technology,\\ 130 Meilong Road, Shanghai 200237, China}

\author{\footnotesize WEI-XING ZHOU}
\address{Department of Finance, Department of Mathematics\\ and Research Center for Econophysics,\\ East China University of Science and Technology,\\ 130 Meilong Road, Shanghai 200237, China\\
wxzhou@ecust.edu.cn}


\maketitle

\begin{history}
\received{26 May 2015}
\end{history}

\begin{abstract}
  China's stock market is the largest emerging market all over the world. It is widely accepted that the Chinese stock market is far from efficiency and it possesses possible linear and nonlinear dependence. We study the predictability of returns in the Chinese stock market by employing the wild bootstrap automatic variance ratio test and the generalized spectral test. We find that the return predictability vary over time and significant return predictability is observed around market turmoils. Our findings are consistent with the Adaptive Markets Hypothesis and have practical implications for market participants.
\end{abstract}


\section{Introduction}
\label{S1:Intro}

The Efficient Markets Hypothesis (EMH) is one of the cornerstones of modern finance.\cite{Fama-1970-JF,Fama-1991-JF} There are three forms of market efficiency, including weak-form, semi-strong form and strong form. The weak-form market efficiency hypothesis suggests the failure of detecting mispriced assets and furthermore the futility of return prediction in terms of related past information. The EMH stimulated numerous studies reporting ``abnormal'' phenomena that are inconsistent with the EMH. One abnormal phenomenon is related to return predictability in terms of historical firm-specific information, such as market capitalization or the size effect,\cite{Banz-1981-JFE} price-earnings ratio,\cite{Basu-1977-JF} book-to-market ratio or the value effect,\cite{Rosenberg-Reid-Lanstein-1985-JPM,Asness-Moskowitz-Pedersen-2013-JF} past prices or the momentum/contrarian effect,\cite{DeBondt-Thaler-1985-JF,Jegadeesh-Titman-1993-JF} and so on. Other anomalies concern with the abnormal returns associated with some calendar times, also called calendar effects, including the weekend effect,\cite{French-1980-JFE} the day of the week effect,\cite{Gibbons-Hess-1981-JB} the January effect,\cite{Sias-Starks-1997-JF} the turn of month effect,\cite{Ariel-1987-JFE} and several others.\cite{Kolb-Rodriguez-1987-JF,Lakonishok-Smidt-1988-RFS,Ariel-1990-JF} These market anomalies cannot be explained by the Capital Asset Pricing Model (CAPM).

The literature about market anomalies has grown in the past three decades and most empirical evidence indicates the deviation of financial markets from the weak-form efficiency. The time-varying performance of market anomalies coincide with the rise and fall of market efficiency. The significant anomalies correspond to the lower level of market efficiency and the weakened anomalies represent higher level of market efficiency. For instance, Urquhart and McGroarty found that the time-varying adjusted returns are related to some calendar effects.\cite{Urquhart-McGroarty-2014-IRFA} They took the data of DJIA as an sample to study different kinds of calendar effects and their findings are in accordance with the implications of Lo's Adaptive Markets Hypothesis (AMH).\cite{Lo-2004-JPM} They also found that some calendar effects are statistically significant merely during certain market conditions, such that the return predictability is time-varying.

A rich variety of methods are employed to conduct research on return predictability through checking the serial correlation of return series, including the methods of Martingale Difference Hypothesis (MDH) test and long-range dependence check.
The MDH testing methods are to verify the existence of linear and nonlinear serial correlation. In this vein, moving windows are usually adopted in most studies to depict the detailed dynamics of return predictability. First, there are numerous methods suitable for investigating possible linear dependence of return time series,\cite{Kim-Shamsuddin-Lim-2011-JEF,Urquhart-Hudson-2013-IRFA,Hiremath-Kumari-2014-SpringerPlus} such as
the portmanteau Q statistic,\cite{Ljung-Box-1978-Biometrika} the AR model,\cite{Ito-Sugiyama-2009-EL} the runs test,\cite{Cohen-Menjoge-1988-JSPI} and the variance ratio test.\cite{Lo-MacKinlay-1988-RFS} Second, there are also method for testing nonlinear dependence in return time series,\cite{Lim-Brooks-Kim-2008-IRFA,Kim-Shamsuddin-Lim-2011-JEF,Urquhart-Hudson-2013-IRFA,Lim-Hooy-2013-TMS,Hiremath-Kumari-2014-SpringerPlus} such as the test for ARCH effect,\cite{McLeod-Li-1983-JTSA} Tsay's test,\cite{Tsay-1986-Bm} the ARCH Lagrange multiplier test,\cite{Engle-1982-Em} the bicorrelation test,\cite{Hinich-1996-JNS} Broock, Scheinkman, Dechert and LeBaron's test,\cite{Broock-Scheinkman-Dechert-LeBaron-1996-EmR} and the generalized spectral test.\cite{Escanciano-Velasco-2006-JEm}
Third, there are econophysical methods for the test of long-range dependence by estimating the Hurst index,\cite{Hurst-1951-TASCE} such as the rescaled range analysis\cite{Mandelbrot-Wallis-1969b-WRR,Mandelbrot-1970-Em,Mandelbrot-1997} the structure function approach,\cite{DiMatteo-Aste-Dacorogna-2005-JBF,DiMatteo-2007-QF} the detrending moving average analysis,\cite{Alessio-Carbone-Castelli-Frappietro-2002-EPJB,Carbone-2009-IEEE,Gu-Zhou-2010-PRE} the detrended fluctuation analysis,\cite{Kantelhardt-KoscielnyBunde-Rego-Havlin-Bunde-2001-PA,Jiang-Xie-Zhou-2014-PA} and so forth.

In this work, we focus on the Chinese stock market. Compared with mature financial markets, the Chinese stock market is relatively young. Some unique phenomena characterize the Chinese market, such as the proportion of retail and institution traders, the universal concept of irrational investment, the impact of policy events, etc. We wonder whether the Chinese market is gradually improving to be efficient or evolving in the way similar to the descriptions in the AMH. In addition, few researches on the AMH have paid attention to the Chinese market. Our study aims to provide more empirical evidence of Chinese market to the literature on the AMH.
We carry out our study from different angles. The classic methods for testing the MDH, including the methods for testing linear and nonlinear dependence of return series, are employed to check the return predictability.

The rest of this paper is organized as follows. Section \ref{S1:Methods} describes the methods employed in this work. Section \ref{S1:Data} presents the data in the study. Section \ref{S1:Results} reports the empirical results. Section \ref{S1:Conclusion} concludes.

\section{Methodology}
\label{S1:Methods}


The EMH states that asset prices fully reflect all available information, which implies the failure of technical analyses. Various methods for testing Martingale difference Hypothesis (MDH) are frequently used by studying return predictability. Specifically, it is a classic assumption in the EMH that
\begin{equation}
\centering
   E[Y_{t}|I_{t-1}]=0,
   \label{Eq:EMH}
\end{equation}
where $\{Y_{t}\}_{-\infty}^{+\infty}$ is a stationary time series, $I(t)$ is the set of all available information before time $t$, and $I_{t} = \{Y_{t}, Y_{t-1}, ...\}$. According to the Eq.~(\ref{Eq:EMH}), $Y_{t}$ is a martingale difference sequence (MDS). The MDH generalizes the notion of MDS such that the unconditional mean of $Y_{t}$ could be nonzero:
\begin{equation}
\centering
   E[Y_{t}|I_{t-1}] = \mu \neq 0,
   \label{Eq:EMH:v1}
\end{equation}
According to equation (\ref{Eq:EMH:v1}), the conditional expectation of $Y_{t}$ is a constant. The MDH implies that conditional mean is independent, which is consistent with the EMH. It means that historical information is useless in forecasting future values. Eq.~(\ref{Eq:EMH:v1}) can be rewritten as follows,
\begin{equation}
\centering
   E[(Y_{t}-\mu)\omega (I_{t-1})]=0,
   \label{Eq:EMH:v2}
\end{equation}
where $\omega(I_{t-1})$ represents the transformation of past information. Different function forms of $\omega$ lead to different methods for testing linear and nonlinear dependence in the return time series.


Charles {\emph{et al.}} compared different MDH testing methods through conducting Monte Carlo experiments.\cite{Charles-Darn-Kim-2011-EL} They concluded that the wild bootstrap automatic variance ratio test (hereafter, AVR test)\cite{Kim-2009-FRL} and the generalized spectral test (hereafter, GS test)\cite{Escanciano-Velasco-2006-JEm} are more favorable to test the linear and nonlinear dependence in return time series. Hence, we use both the AVR test and the GS test to investigate the return predictability in the Chinese stock market. We review briefly these two methods.

\subsection{Wild bootstrap automatic variance ratio test}

The variance ratio test was developed by Lo and MacKinlay,\cite{Lo-MacKinlay-1988-RFS} which has been widely employed to test if a market is efficient in the weak form. Let $Y_{t}$ be the asset return at time $t$ ($t=1,...,T$), the AVR test statistic can be written as follows:\cite{Choi-1999-JAE}
\begin{equation}
\centering
   VR(k)=1+2\sum_{i=1}^{T-1}m\left({i}/{k}\right)\hat{\rho}(i) =1+2\sum_{i=1}^{T-1}m\left({i}/{k}\right)\frac{\sum_{t=1}^{T-i}(Y_t-\hat{\mu})(Y_{t+i}-\hat{\mu})}{\sum_{t=1}^{T}(Y_t-\hat{\mu})^2}
   \label{Eq:VR1}
\end{equation}
where $k$ is the holding period, $\hat{\rho}(i)$ is the estimator for $i$th order autocorrelation of the returns, $\hat{\mu}=\frac{1}{T}\sum_{t=1}^{T}Y_t$, and the quadratic spectral kernel $m(x)$ is the weighting function
\begin{equation}
\centering
   m(x)=\frac{25}{12{\pi}^{2}x^{2}}\left[\frac{\sin\left(\frac{6{\pi}x}{5}\right)}{\left(\frac{6{\pi}x}{5}\right)}
   -\cos\left(\frac{6{\pi}x}{5}\right)\right]
   \label{Eq:VR:Kernel}
\end{equation}
When $Y_t$ is i.i.d. and has finite fourth moment as well, under the null hypothesis that $Y_t$ is serially uncorrelated, we have\cite{Choi-1999-JAE}
\begin{equation}
\centering
   AVR(k)=\sqrt{\frac{T}{k}}\frac{[VR(k)-1]}{\sqrt{2}}\xrightarrow[]{d}N(0,1)
   \label{Eq:VR:AVR}
\end{equation}
as $k \rightarrow \infty$, $T \rightarrow \infty$, $T/k \rightarrow \infty$. The optimal value of lag truncation point (or holding period) $k$ can be determined by the fully data-dependent method.\cite{Andrews-1991-Em} One thus obtains the AVR test statistic $AVR(\hat{k})$ with the optimal choice $\hat{k}$ for $k$.

It is argued that the small sample properties of the AVR test\cite{Choi-1999-JAE} can be substantially improved after employing the wild bootstrap. Extensive Monte Carlo experiments have been conducted to show that the wild bootstrap provides accurate statistical inference in small samples under conditional heteroskedasticity.\cite{Kim-2006-EL,Kim-2009-FRL} Specifically, wild bootstrap of $AVR(\hat{k})$ could be performed as the following three steps:\cite{Kim-2006-EL,Kim-2009-FRL} (1) Generate a bootstrap sample of size $T$, $Y_{t}^{*}=\eta_{t}Y_{t}$ ($t=1,...,T$), where $\eta_{t}$ is random variable with zero mean and unit variance; (2) Obtain the $AVR^{*}(k^{*})$ through calculating AVR statistic from $\{Y_{t}^{*}\}_{t=1}^{T}$; and (3) Repeat the first two steps many times and construct a bootstrap distribution  $\{AVR^{*}(\hat{k^{*}};j)\}_{j=1}^{B}$.

\subsection{Generalized spectral test}

On the other hand, $\omega$ can be nonlinear functions. Exponential function and indicator function are popularly adopted. The former is to detect the general nonlinear conditional mean dependence, and the latter is to test for no directional predictability.

Escanciano and Velasco proposed the null of the MDH in a form of pairwise regression function.\cite{Escanciano-Velasco-2006-JEm} The null hypothesis is that $H_0$: $m_{j}(y)=0$, where $m_{j}(y)=E(Y_{t}-\mu|Y_{t-j}=y)$, and the alternative hypothesis is that $H_{1}$: $P\{m_{j}(y)\neq 0\}>0$ for some $j$. In fact, the above null hypothesis is consistent with the exponential weighting function as follows,
\begin{equation}
\centering
   \gamma_{j}(x)\equiv E[(Y_{t}-\mu) \exp{(ixY_{t-j})}]=0
   \label{Eq:ExponentialWeight}
\end{equation}
where $\gamma_{j}(x)$ plays a role of an autocovariance measure in a nonlinear framework with $x$ being any real number. They also proposed the use of the generalized spectral distribution function,\cite{Escanciano-Velasco-2006-JEm}
\begin{equation}
\centering
   H(\lambda,x)=\gamma_{0}(x)\lambda+2\sum_{j=1}^{\infty}\gamma_{j}(x)\frac{\sin(j\pi\lambda)}{j\pi}
   \label{Eq:GSDF}
\end{equation}
whose sample estimate is written as
\begin{equation}
\centering
   \hat{H}(\lambda,x)=\hat{\gamma}_{0}(x)\lambda+2\sum_{j=1}^{\infty}\left(1-\frac{j}{T}\right)\hat{\gamma}_{j}(x)
   \frac{\sin(j\pi\lambda)}{j\pi}
   \label{Eq:GSDF2}
\end{equation}
where $\hat{\gamma}_{j}(x)=(T-j)^{-1}\sum_{t=1+j}^{T}(Y_{t}-\bar{Y}_{T-j})e^{ixY_{t-j}}$ and $\bar{Y}_{T-j}=(T-j)^{-1}\sum_{t=1+j}^{T}Y_{t}$. Under the null, $\hat{H}(\lambda,x)=\hat{\gamma}_{0}(x)\lambda \equiv \hat{H}_{0}(\lambda,x)$, and the test statistic for $H_{0}$ is written as
\begin{equation}
\centering
   S_{T}(\lambda,x)=(0.5T)^{1/2}\left[\hat{H}(\lambda,x)-\hat{H}_{0}(\lambda,x)\right]
   \label{Eq:GSDF3}
\end{equation}

To evaluate the value of $S_{T}$ for all possible values of $\lambda$ and $x$, the Cramer-von Mises norm is used to obtain the statistic\cite{Escanciano-Velasco-2006-JEm}
\begin{equation}
\centering
   D_{T}^{2}=\sum_{j=1}^{T-1}\frac{T-j}{(j\pi)^2}\sum_{t=j+1}^{T}\sum_{s=j+1}^{T}
   (Y_{t}-\bar{Y}_{T-j})(Y_{s}-\bar{Y}_{T-j})\exp\left[-0.5(Y_{t-j}-Y_{s-j})^2\right]
   \label{Eq:CMnorm}
\end{equation}
To improve small sample properties, Escanciano and Velasco recommended the use of the wild bootstrap,\cite{Escanciano-Velasco-2006-JEm} whose process is similar to that in the AVR test mentioned above.

\section{Data sets}
\label{S1:Data}

The data used to study the return predictability through the AVR and GS tests are retrieved from RESSET (http://www.resset.cn), which contains the daily and weekly returns for all A-share individual stocks listed on the Shanghai Stock Exchange (SHSE) and the Shenzhen Stock Exchange (SZSE), covering the period from December 1990 to September 2015. The equally-weighted average daily and weekly returns of all individual stocks are calculated for both exchanges. Table \ref{TB:DS:ChinaDaily} presents some related descriptive statistics. It is found that SHSE stocks have higher daily and weekly average returns than SZSE stocks during the sample period, which is consistent with the fact that the average returns of SHSE stocks have larger skewness. In addition, the average returns of SHSE stocks have higher kurtosis than SZSE stocks. The Jarque-Bera test also shows that the averages are not normally distributed.

\begin{table}[!ht]
\centering
\caption{
{Descriptive statistics of weekly and daily average returns for the SHSE and SZSE stocks. The Jarque-Bera statistics are presented in last column and superscripts *** correspond to the significance level of 1\%, which indicates that the series do not follow the normal distribution.}}
   \label{TB:DS:ChinaDaily}
   \begin{tabular}{cccccccc}
   \hline
    & \textit{Market} & \textit{size} & \textit{Mean} & \textit{Std.} & \textit{Skew.} & \textit{Kurt.}  & \textit{Jarque-Bera}\\
   \hline
  Weekly & SHSE&  1254 & 0.00775 & 0.083 & 11.21 & 214.3 &  2359173$^{***~}$ \\
  Weekly & SZSE&  1250 & 0.00509 & 0.061 &  3.24 &  42.4 &    83133$^{***~}$ \\
  Daily  & SHSE&  6063 & 0.00149 & 0.030 & 14.25 & 591.3 & 87635173$^{***~}$ \\
  Daily  & SZSE&  6102 & 0.00099 & 0.025 &  1.23 &  22.2 &    95593$^{***~}$ \\
   \hline
   \end{tabular}
\end{table}

\section{Empirical results}
\label{S1:Results}

We perform the AVR test and the GS test for the return predictability of the daily and weekly data in the SHSE and the SZSE. Because the observations in the corresponding time window are sufficient to guarantee the precise estimation, the size of time window for weekly data and daily data is set as 2 years and $5$ years, respectively.\cite{Kim-Shamsuddin-Lim-2011-JEF} Each time window includes nearly $250$ observations for weekly data and almost $500$ observations for daily data. The sample is moved one year forward for the re-estimation of the AVR and GS statistics.

\begin{figure}[!ht]
  \centering
  \includegraphics[width=6cm]{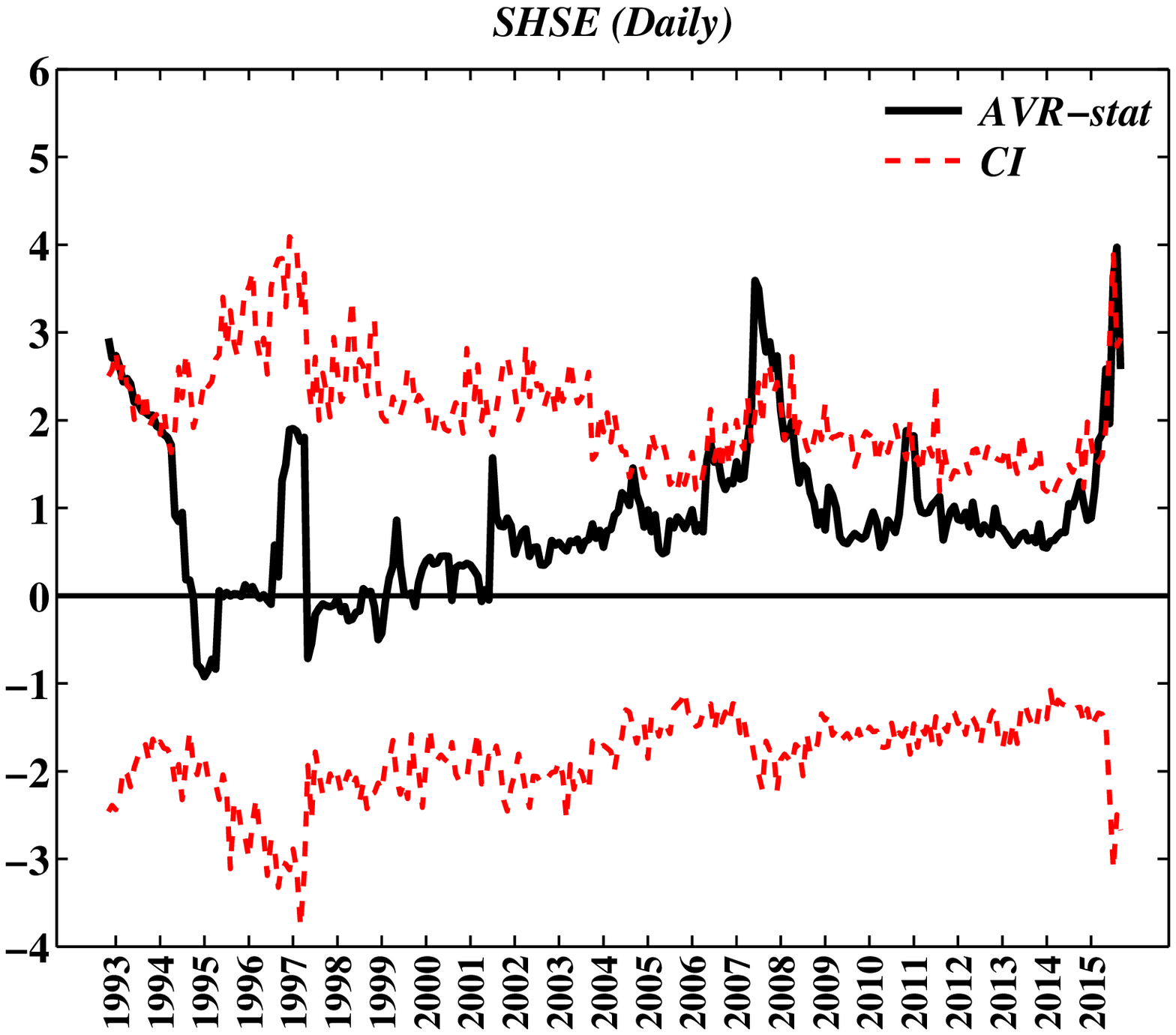}
  \includegraphics[width=6cm]{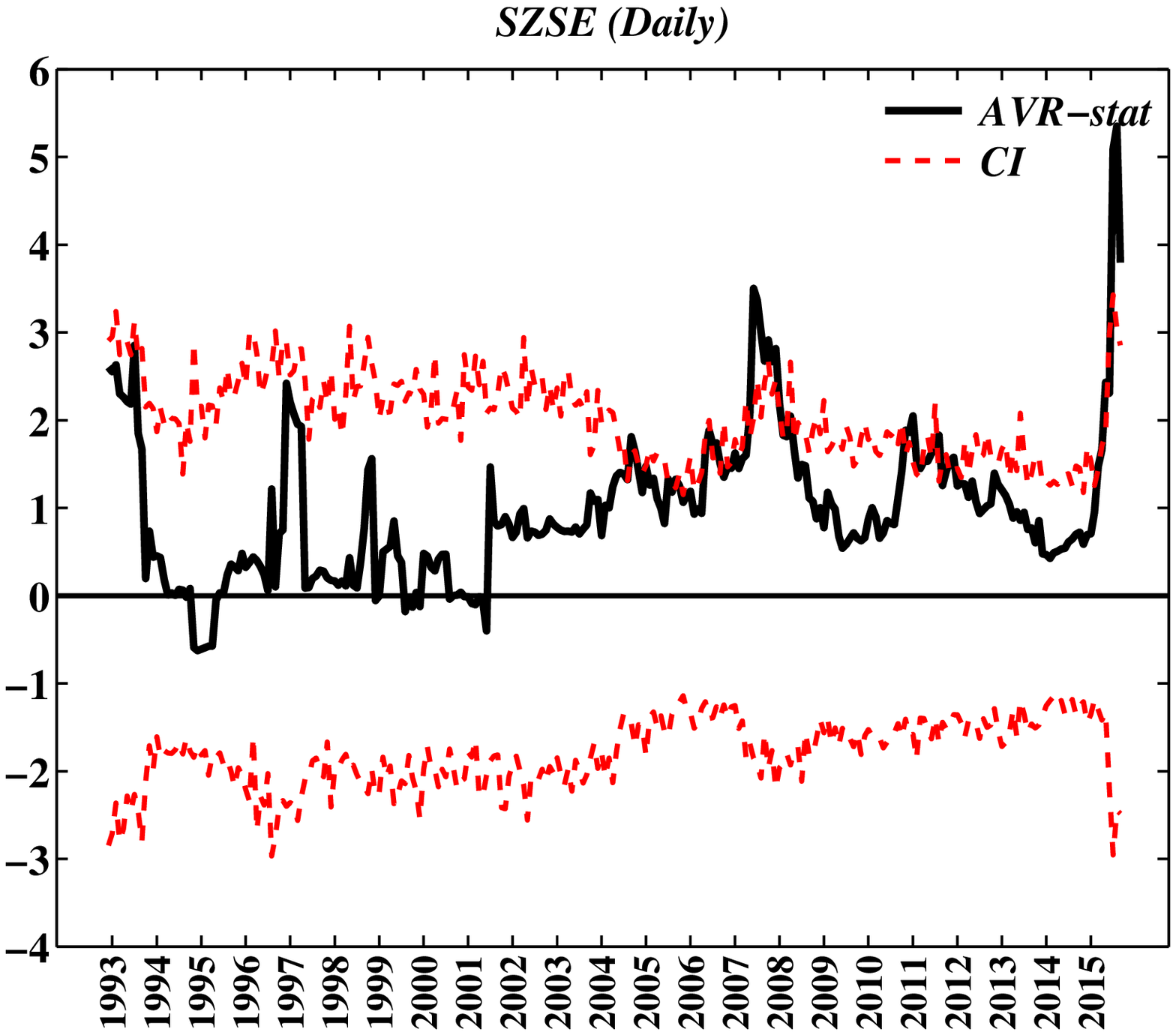}
  \includegraphics[width=6cm]{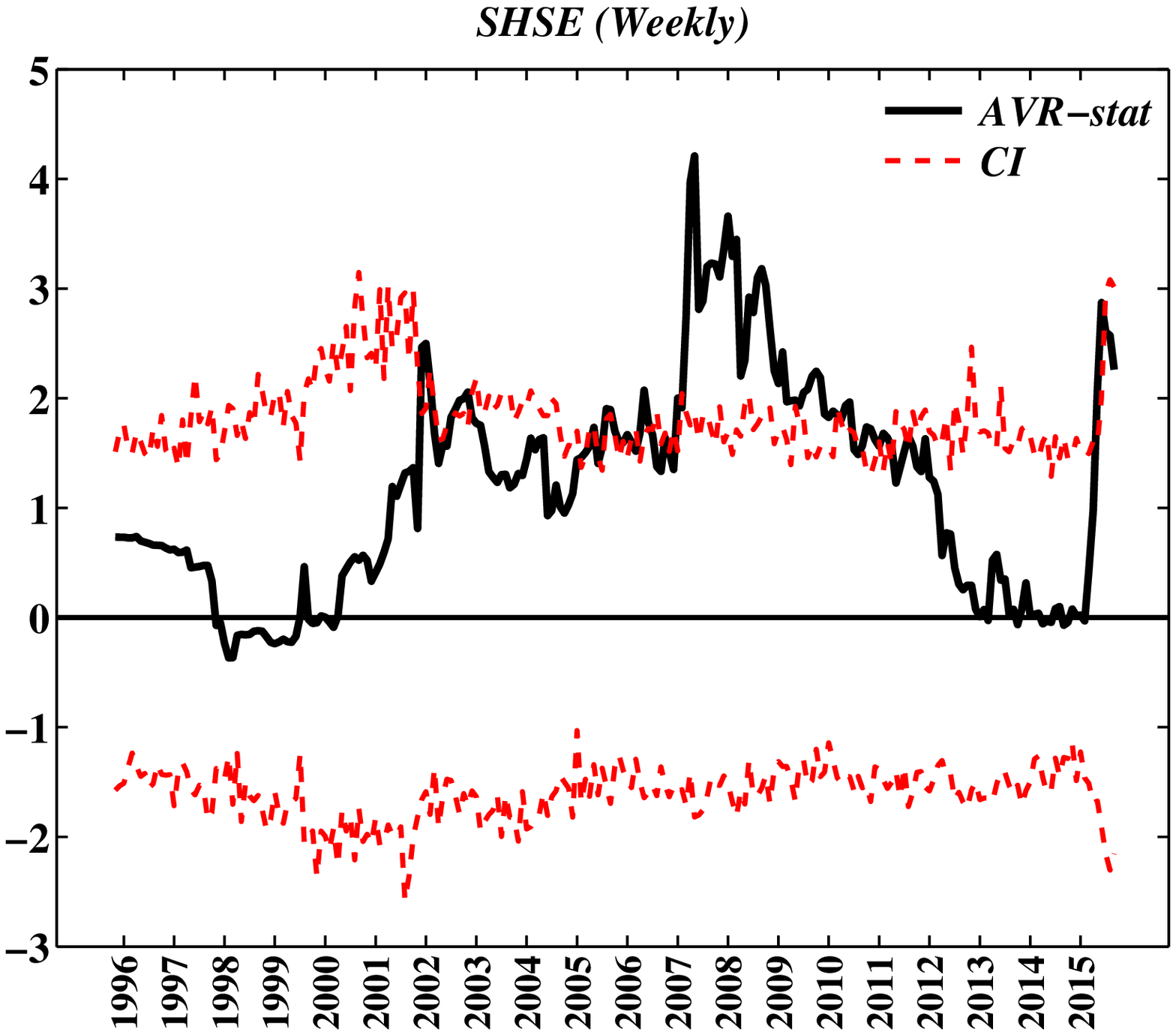}
  \includegraphics[width=6cm]{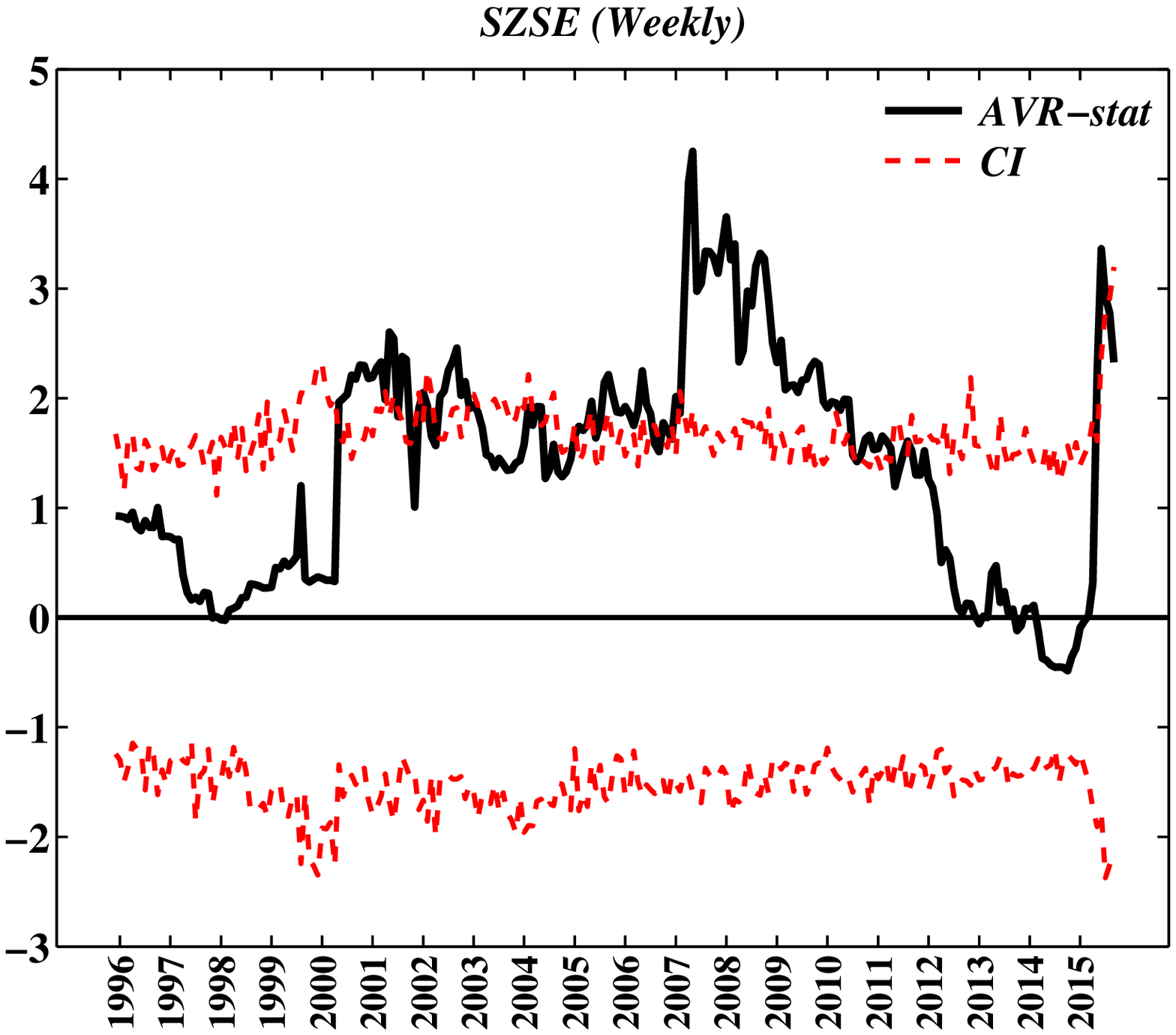}
  \caption{Time-varying statistics of the AVR test of the daily and weekly data in the SHSE and the SZSE from December 1990 to September 2015. CI is the confidence interval associated with the statistic at the significance level of $5\%$. The time $t$ represents the time window $[t-1,t]$ for daily data, and $[t-4,t]$ for weekly data.}
  \label{Fig:Rolling:AVRstat}
\end{figure}

The time-varying AVR statistic and its corresponding 5\% confidence intervals (CIs) are illustrated in Fig.~\ref{Fig:Rolling:AVRstat}. When the AVR statistic is greater than the upper CI value, the returns exhibit statistically significant positive serial correlation. When the AVR statistic is less than the lower CI value, the returns exhibit statistically significant negative serial correlation. It is evident that the AVR statistic rises and falls over time. Most of the AVR statistic values are positive, be them statistically significant or not, suggesting the presence of positive linear correlation in return time series in the Chinese stock market. These results are not sensitive to the two exchanges and data sample frequency.

Kim {\emph{et al.}} found that stock market bubbles and crashes generally correspond to higher AVR statistics and sometimes wider confidence bands, which can serve as the measure of market uncertainty.\cite{Kim-Shamsuddin-Lim-2011-JEF} It implies that the market efficiency is dependent on market conditions.\cite{Lo-2004-JPM,Kim-Shamsuddin-Lim-2011-JEF} Specifically, when the Chinese stock exchanges (including the SHSE and SZSE) were established in the early 1990s, the markets were more volatile and had higher uncertainty mainly due to the market trading mechanism of $T+0$ and no implementation of mature pricing limits. Accordingly, for daily data in the early 1990s, the AVR statistics are positive with statistical significance and the confidence band is wider. Similarly, during the time period from 1996 to 1997, the results for daily data show that the AVR statistics have a sudden and sharp increase with the CI band getting wider, which corresponds to the market bubble in Chinese stock market from 1996 to 1997. During the time period around 2001, the results for both daily and weekly data show a rapid increase of the AVR statistics and the corresponding confidence interval band, representing an increase of market uncertainty. On 2001/06/12, the policy of state-owned shares reduction was released by the State Council of China, which triggered an long-lasting antibubble.\cite{Zhou-Sornette-2004a-PA} The SHSE Composite Index culminated at the all-time high of 2242.4 on 2001/06/13 and since plummeted 32.2\% to 1520.7 on 2001/10/22 when the policy was ceased. The most intriguing patterns of the results in Fig.~\ref{Fig:Rolling:AVRstat} appear in 2007 and 2015, corresponding to the two infamous crashes following two huge bubbles from 2005 to December 2007\cite{Jiang-Zhou-Sornette-Woodard-Bastiaensen-Cauwels-2010-JEBO} and from 2014 to June 2015.\cite{Sornette-Demos-Zhang-Cauwels-Filimonov-Zhang-2015-JIS} It is found that the AVR statistic is significantly positive around the crash, indicating a very different market correlation structure.\cite{Han-Xie-Xiong-Zhang-Zhou-2017-FNL}

\begin{figure}[!ht]
  \centering
  \includegraphics[width=6cm]{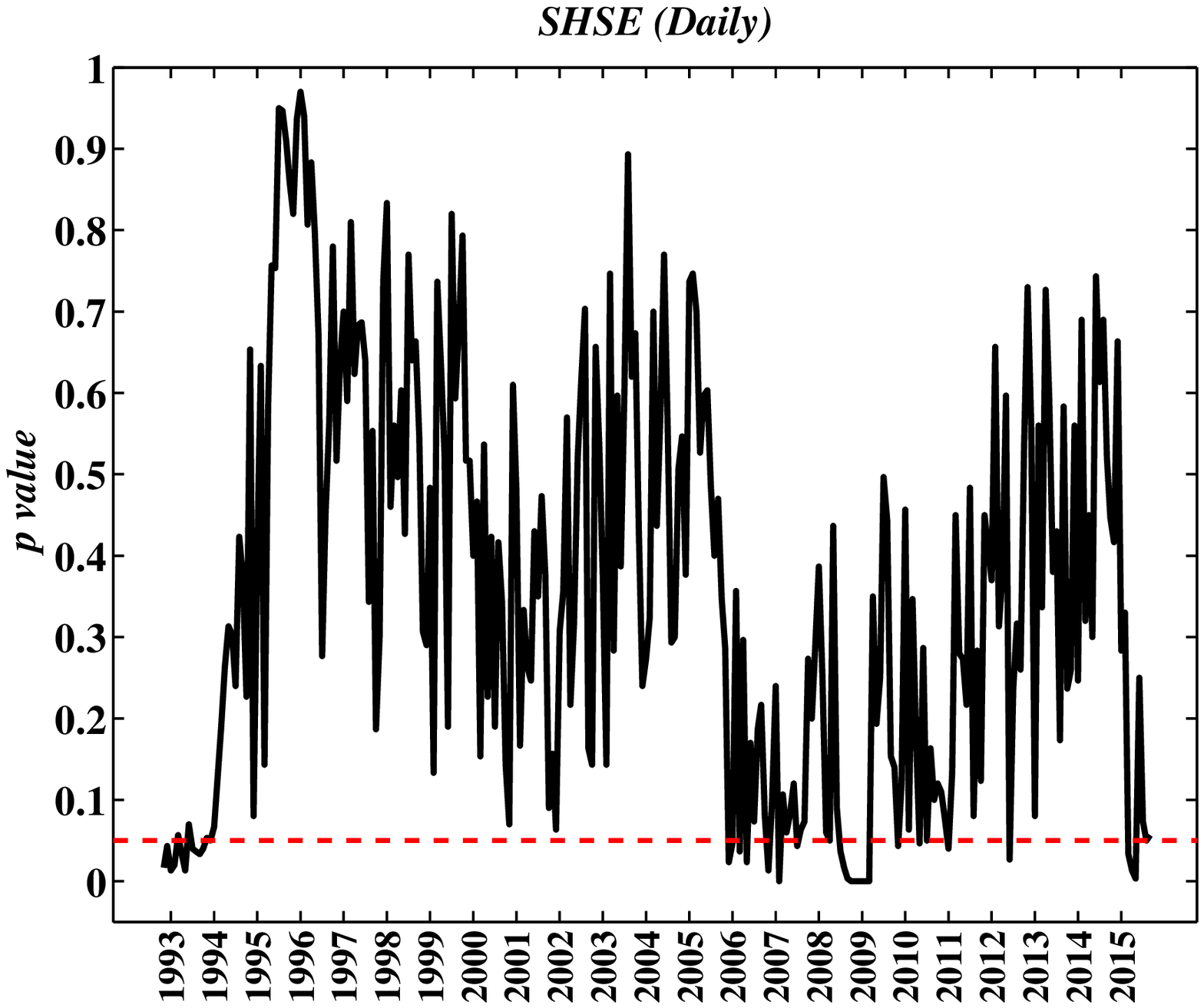}
  \includegraphics[width=6cm]{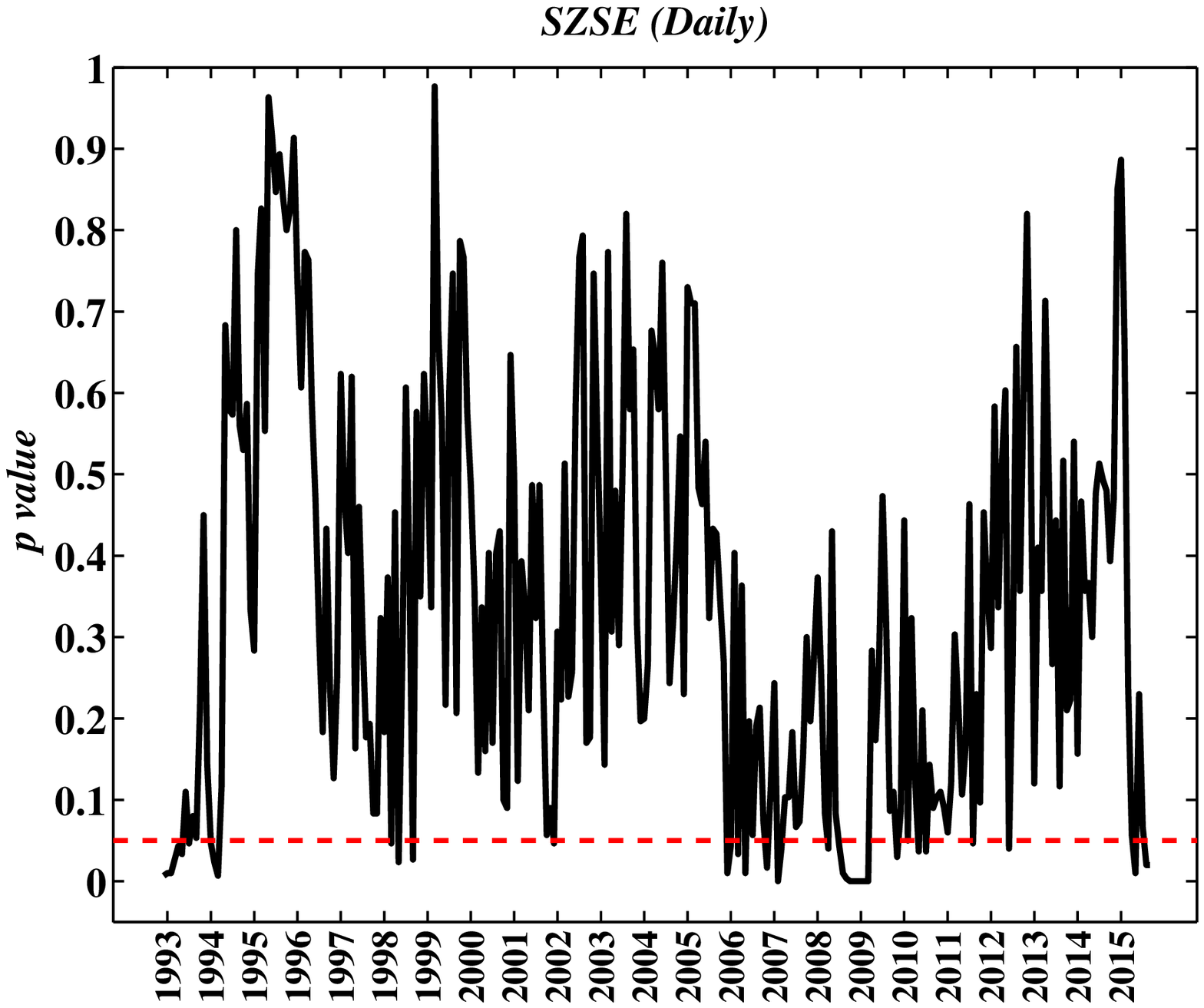}
  \includegraphics[width=6cm]{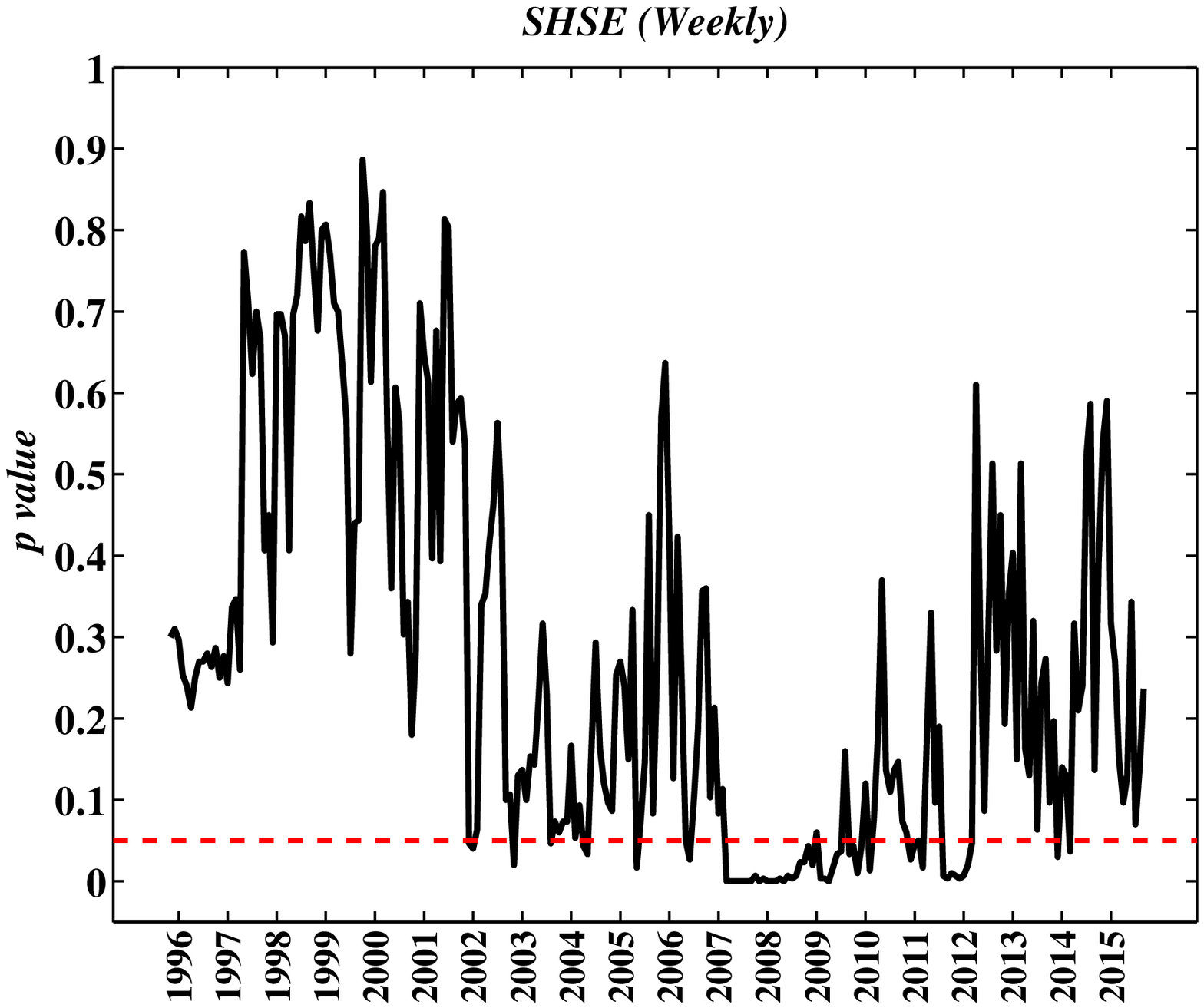}
  \includegraphics[width=6cm]{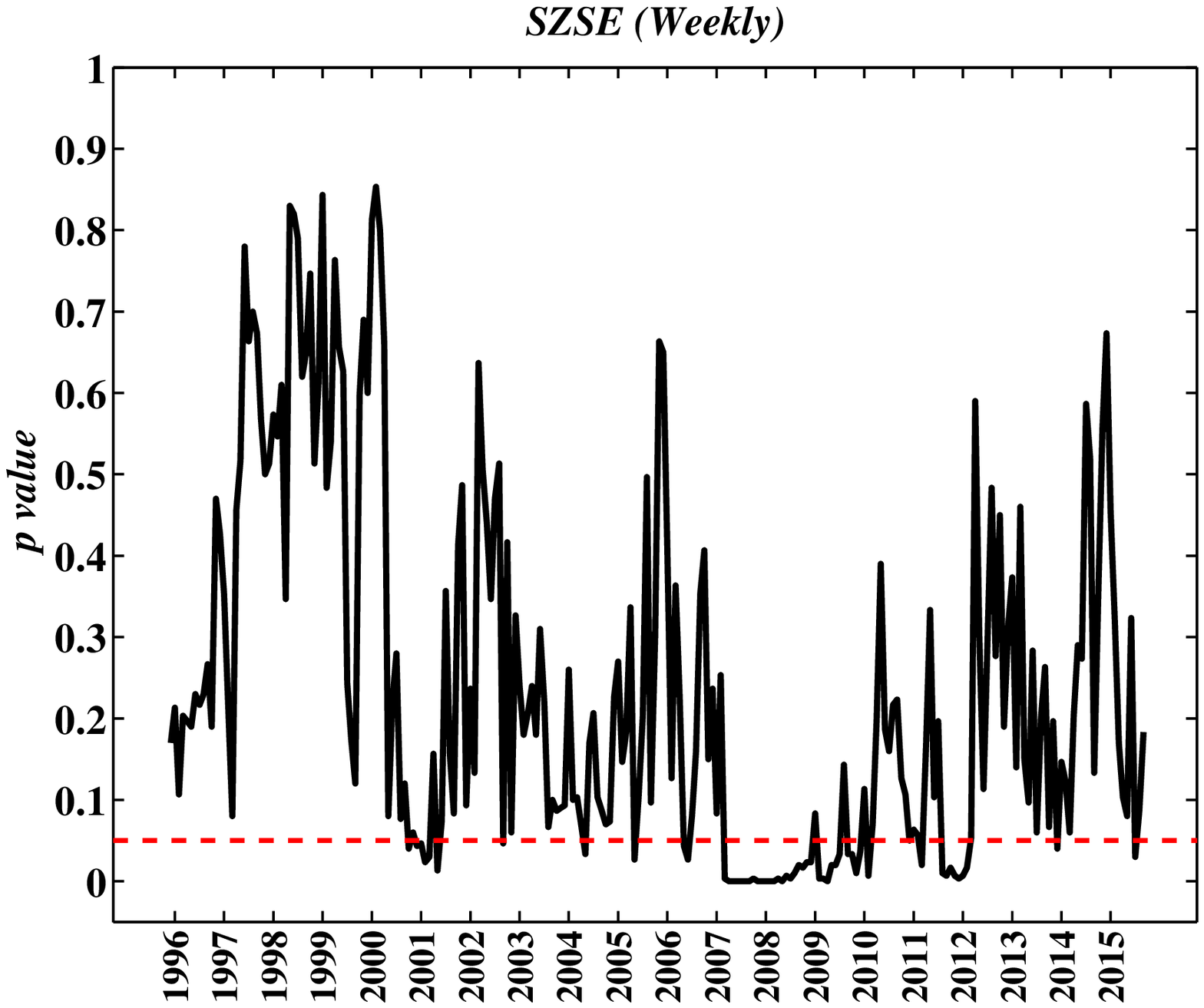}
  \caption{\label{Fig:Rolling:GSpval} Time-varying $p$-values for the GS test statistics of the daily and weekly data in the SHSE and the SZSE from December 1990 to September 2015. The horizontal dashed lines correspond to the $0.05$ significance level. The time $t$ represents the time window $[t-1,t]$ for daily data, and $[t-4,t]$ for weekly data.}
\end{figure}

Fig.~\ref{Fig:Rolling:GSpval} depicts time-varying $p$-values for the GS statistics. The return time series exhibit statistically significant nonlinear dependence if the $p$-value falls down beneath the dashed line corresponding to the significance level of 5\%. We observe that the $p$-values fluctuate over time. Significant nonlinear dependence appears in several time periods, similar to the results of the AVR test. The most evident is again around the 2007 crash. However, it seems that the AVR statistic has stronger predictive power of large crashes. Rigorous evaluation of the predictive power can be carried out based on the pattern recognition framework.\cite{Sornette-Zhou-2006-IJF}

Apparently, the AVR and GS statistics, as measures of market efficiency, indicate that the efficiency of the Chinese stock market is time-varying and dependent on market conditions. Specifically, there would be higher degree of return predictability during market bubbles and crashes, which is different from the findings in US market.\cite{Kim-Shamsuddin-Lim-2011-JEF} All these findings are in accordance with the Adaptive Markets Hypothesis.\cite{Lo-2004-JPM,Lo-2005-JIC}

\section{Conclusion}
\label{S1:Conclusion}

We have adopted the wild bootstrap automatic variance ratio test and the generalized spectral test to check the return predictability in the Chinese market. The results show taht the return predictability is time-varying and dependent on market conditions. More specifically, the high levels of return predictability coincide with historical market turmoils around crashes. In some cases, there are clear evidence showing that cumulating return predictability is able to predict market crashes. 

Our findings indicate that the Chinese stock market is inefficient and the level of market inefficiency varies over time. These results provide evidence supporting the Adaptive Markets Hypothesis.

\section*{Acknowledgments}

This work was partly supported by the National Natural Science Foundation of China (Grants 71131007, 71532009) and the Fundamental Research Funds for the Central Universities.

\section*{References}

%

\end{document}